\documentstyle[amssymb,preprint,aps]{revtex}

\newtheorem{theorem}{Theorem}
\newtheorem{acknowledgement}[theorem]{Acknowledgement}

\begin{document}
\title{Spontaneous Transitions in Quantum Mechanics}
\author{E. Prodan}
\address{University of Houston, Department of Physics\\
4800 Calhoun Road, Houston, TX, 77204-5506}
\maketitle

\begin{abstract}
\noindent \noindent \noindent \tightenlines

\noindent The problem of spontaneous pair creation in static external fields
is reconsidered. A weak version of the conjecture proposed in \cite{NE1} is
stated and proved. The method reduces the proof of the general conjecture to
the study of the evolution associated with the time dependent Hamiltonian, $%
H_{\varepsilon }\left( t\right) $, of a vector which is eigenvector of $%
H_{\varepsilon }\left( t\right) $ at some given time. A possible way of
proving the general conjecture is discussed.
\end{abstract}

\section{Introduction}

\noindent \tightenlines

We reconsider in this paper the problem of spontaneous pair creations in
static external fields. In the original version \cite{NE1}, the problem was
addressing to high energy physicists. The experimental test was done by
comparing the theoretical predictions with the experimental results coming
from heavy ion collision experiments. As is stated in \cite{NE2}, there was
no agreement between the two results, one of the possible cause being the
large effects of non-adiabatic processes.

\noindent In the past few years, experimental results showed that the
transport properties of the semiconductors with high symmetry may change
drastically if a certain critical value of the external electric field is
exceed. A particular example is a quasi-one dimensional semiconductor,
cooled down below the Peierls transition temperature. It is known that,
below this critical temperature, a gap is opening in the single-particle
excitation spectrum. Moreover, the experimental results \cite{GO} show the
existence of a threshold value of the applied electric field where the
transport properties change drastically. The two elements: existence of the
gap in the one-particle Hamiltonian spectrum and the existence of the
critical value of the applied electric field, above which the conductivity
is practically reduced to zero, are strong arguments for the idea that we
are facing here with the phenomenon of spontaneous pair creations. We agree
that there are many theories which, more or less, explain this phenomenon.
While most of them involve interacting quantum fields, our hope is that an
effective potential can be written down such that, for applied electric
fields above the threshold value, the overcritical part of the conjecture 
\cite{NE1} applies. If this is true, then there may be another way to
experimentally test the theory, this time, with a better control on the time
variations of the external fields and so, on the non-adiabatic processes. In
some situations, the threshold value of the electric field can be small.
This means that, experimentally, we are not enforced to switch off the
applied field (to protect the sample). This shows one of the qualitative
difference between the two experimental settings: in the heavy ion
collisions, the quantum system is perturbed by the electric fields produced
during the collisions so we have no control on the ``switch on'' or ``switch
off'' of the interaction. In contradistinction, for a semiconductor with low
critical value of the electric field, we have total control on how slow the
interaction is introduced.

\noindent Because of a technical difficulty, in \cite{NE2}, the definition
of overcritical external fields was slightly modified in order to prove the
existence of the overcritical external fields. We propose another approach
of the problem which avoid this technical difficulty. However, this doesn't
mean that the problem of spontaneous pair creation is solved, but, in the
light of the last observation, the new approach seems to be more appropriate
for the problem of spontaneous pair creations in semiconductors.

\section{Description of the problem}

\noindent Because the results in scattering problems involving periodic
Schrodinger operators are much poor than for those involving Dirac
operators, we will treat the problem at the level of first quantization. We
show that, above the critical value of the interaction, electrons can
spontaneously transit between two different energetic bands. If the
scattering operator can be implemented in the second quantization, this
result is equivalent with spontaneous pair creations of electrons and holes.

\noindent For simplicity, we will discuss here the case of a self-adjoint
operator, $H_{0}$, defined on some dense subspace ${\cal D}\left(
H_{0}\right) $ of the Hilbert space ${\cal H}$, of whom spectrum consists of
two absolute continuous, bounded, disjoint parts. We denote the lower and
upper parts by $\sigma _{-}$ and $\sigma _{+}$ respectively. Let $H_{\lambda
}=H_{0}+\lambda V$ be the perturbed operator, where we assume that ${\cal D}%
\left( H_{\lambda }\right) ={\cal D}\left( H_{0}\right) $, ${\cal D}\left(
H_{0}\right) \subset {\cal D}\left( V\right) $ and the perturbation leaves $%
\sigma _{-}$ and $\sigma _{+}$ unchanged. Our interest is in the case when,
as $\lambda $ increases, some eigenvalues emerge from $\sigma _{+}$ and move
continuously to $\sigma _{-}$, and there is a critical value, $\lambda _{c}$%
, at which the lowest eigenvalue touches $\sigma _{-}$ and then it
disappears in the lower continuum spectrum. We study the scattering problem
of pair $\left( H_{0},\ H_{\lambda }\right) $ in the adiabatic switching
formalism for both cases: $\lambda <\lambda _{c}$ and $\lambda >\lambda _{c}$%
.

\noindent Let us consider a function, $\varphi :\ {\Bbb R}\rightarrow {\Bbb R%
}_{+}$, $\varphi \in {\cal C}^{\infty }$ such that 
\begin{equation}
\varphi \left( s\right) =\left\{ 
\begin{array}{c}
1\text{, }\left| s\right| <1 \\ 
0\text{, }\left| s\right| >2
\end{array}
\right. 
\end{equation}
and, for a pair of positive numbers, $\varepsilon =\left( \varepsilon
_{1},\varepsilon _{2}\right) $, we consider the adiabatic switching factor: 
\begin{equation}
\varphi _{\varepsilon }\left( s\right) =\left\{ 
\begin{array}{c}
\varphi \left( \varepsilon _{1}s\right) \text{, }s<0 \\ 
\varphi \left( \varepsilon _{2}s\right) \text{, }s\geq 0
\end{array}
\right. .
\end{equation}
One can consider that $\varepsilon _{1}$ controls the ``switch on'' process
and $\varepsilon _{2}$ controls the ``switch off'' process. Note that $%
\varphi _{\varepsilon }$ is also of ${\cal C}^{\infty }$. For the time
dependent Hamiltonian: $H_{\varepsilon ,\lambda }\left( t\right)
=H_{0}+\lambda \varphi _{\varepsilon }\left( t\right) V$, and time
independent Hamiltonian: $H_{\lambda }=H_{0}+\lambda V$, we denote by: 
\begin{equation}
W_{\varepsilon ,\lambda }^{\pm }=s-%
\mathrel{\mathop{\lim }\limits_{T\rightarrow \pm \infty }}%
U_{\varepsilon ,\lambda }^{\ast }\left( T,0\right) e^{-iTH_{0}}
\end{equation}
and 
\begin{equation}
W_{\lambda }^{\pm }=s-%
\mathrel{\mathop{\lim }\limits_{T\rightarrow \pm \infty }}%
e^{iTH_{\lambda }}e^{-iTH_{0}}
\end{equation}
the adiabatic and static Moller operators. The notation $U_{\varepsilon
,\lambda }\left( T,T^{\prime }\right) $ stands for the propagator
corresponding to $H_{\varepsilon ,\lambda }\left( t\right) $. We suppose
that, for $\lambda \in \left[ 0,\lambda _{0}\right] $, $\lambda _{0}>\lambda
_{c}$, these operators exist, the adiabatic Moller operators converge
strongly to the static operators. In addition, we consider that the static
Moller operators are locally complete on $\sigma _{-}$, i.e. $Range\left[
P_{H_{\lambda }}\left( \sigma _{-}\right) W_{\lambda }^{\pm }\right]
=P_{H_{\lambda }}\left( \sigma _{-}\right) {\cal H}$. We will discuss later
why the situation is different in the case when the Moller operators are
only weakly complete (in the sense of \cite{PE}). With these assumptions,
one can define the unitary scattering matrix $S_{\lambda }=\left( W_{\lambda
}^{-}\right) ^{\dagger }\times W_{\lambda }^{+}$ and the adiabatic version, $%
S_{\varepsilon ,\lambda }=\left( W_{\varepsilon ,\lambda }^{\,-}\right)
^{\dagger }\times W_{\varepsilon ,\lambda }^{\,+}$. It is known \cite{DO}
that the adiabatic scattering operator converge weakly to the static
scattering operator in the adiabatic limit, $\varepsilon \rightarrow 0$.

\noindent Let us denote by $P_{H_{\lambda }}\left( \Omega \right) $ the
spectral projection of $H_{\lambda }$ corresponding to some $\Omega \subset R
$. The spontaneous excitations (transfer from $P_{H_{0}}\left( \sigma
_{-}\right) $ to $P_{H_{0}}\left( \sigma _{+}\right) $ and vice-versa) are
denied by the fact that the scattering matrix $S_{\lambda }$ commutes with
the unperturbed Hamiltonian and in consequence: $P_{H_{0}}\left( \sigma
_{\pm }\right) \,S_{\lambda }\,P_{H_{0}}\left( \sigma _{\mp }\right) \equiv 0
$. The key observation is that $S_{\varepsilon ,\lambda }$ does not commute
with the unperturbed Hamiltonian and, because $S_{\varepsilon ,\lambda }$
goes weakly to the static scattering operator, we still have a chance for $%
\lim\limits_{\varepsilon \rightarrow 0}\left\| P_{H_{0}}\left( \sigma _{\pm
}\right) \,S_{\lambda }\,P_{H_{0}}\left( \sigma _{\mp }\right) \right\| >0$.
Indeed, was proven in \cite{NE2} that this is the case if one considers a
discontinuous switching factor, $\varphi _{\delta }$, with $%
\lim\limits_{\delta \rightarrow 0}\varphi _{\delta }$ a smooth function.
Moreover, it was shown that 
\begin{equation}
\lim\limits_{\varepsilon _{1}=\varepsilon _{2}\rightarrow 0}\left\|
P_{H_{0}}\left( \sigma _{\pm }\right) \,S_{\varepsilon ,\lambda
}\,P_{H_{0}}\left( \sigma _{\mp }\right) \right\| =1-o\left( \delta \right) 
\end{equation}
provided $\lambda >\lambda _{c}$. We will prove in the next section that: 
\begin{equation}
\lim\limits_{\varepsilon _{1}\rightarrow 0}\lim\limits_{\varepsilon
_{2}\rightarrow 0}\left\| P_{H_{0}}\left( \sigma _{\pm }\right)
\,S_{\varepsilon ,\lambda >\lambda _{c}}\,P_{H_{0}}\left( \sigma _{\mp
}\right) \right\| =1\text{,}
\end{equation}
but with $\varphi $ of ${\cal C}^{\infty }$ class. As was already pointed
out in the previous section, this version may be more appropriate for the
case of pair creations in semiconductors.

\section{The result}

\noindent Our main result is:

\begin{theorem}
In the conditions enunciated in the previous sections, for $\lambda \in %
\left[ 0,\lambda _{0}>\lambda _{c}\right] $ and $H\left( t\right) $ of $%
{\cal C}^{3}$ in respect with $t$ (in the sense of \cite{NE3}), then: 
\begin{equation}
\lim\limits_{\varepsilon _{1}\rightarrow 0}\lim\limits_{\varepsilon
_{2}\rightarrow 0}\left\| P_{H_{0}}\left( \sigma _{-}\right) S_{\varepsilon
,\lambda }P_{H_{0}}\left( \sigma _{+}\right) \right\| =\left\{ 
\begin{array}{c}
0\text{ if }\lambda <\lambda _{c} \\ 
1\text{ if }\lambda >\lambda _{c}
\end{array}
\right. \text{.}
\end{equation}
\end{theorem}

\noindent {\bf Proof. }The under-critical part ($\lambda <\lambda _{c}$)
results directly from the adiabatic theorem. In this situation, the order of
limits are non-important. Note that the under-critical case was proven in
full generality for Dirac operators in \cite{NE1}.\newline
We start now the proof of the overcritical part ($\lambda >\lambda _{c}$)
which follows closely \cite{NE2}. We will denote by $E_{g}\left( t\right) $
and $\psi _{g}\left( t\right) $ the lowest eigenvalue of $H_{\varepsilon
,\lambda }\left( t\right) $ and one of its eigenvector. (Without loss of
generality, we can suppose that the eigenvalues do not change their order
during the switching). Any constant which depends on $\varepsilon _{1,2}$
and goes to zero as $\varepsilon _{1,2}$ goes to zero will be denoted by $%
o\left( \varepsilon _{1,2}\right) $. Our task is to find a vector $\phi $, $%
\left\| \phi \right\| =1$, such that 
\begin{equation}
\left\| P_{H_{0}}\left( \sigma _{-}\right) S_{\varepsilon ,\lambda
}P_{H_{0}}\left( \sigma _{+}\right) \phi _{\varepsilon }\right\| >1-o\left(
\varepsilon _{1},\varepsilon _{2}\right) \text{.}
\end{equation}
Let $\varphi \left( -s_{0}\right) =\lambda _{c}/\lambda $, $s_{0}>0$, and $%
0<\delta <1$ such that $E_{g}\left( -\left( s_{0}+\delta \right)
/\varepsilon _{1}\right) $ exists. From the adiabatic theorem applied on $%
\left( -2/\varepsilon _{2},-\left( s_{0}+\delta \right) /\varepsilon
_{1}\right) $ we get 
\begin{equation}
\left\| P_{H_{0}}\left( \sigma _{+}\right) U_{\varepsilon ,\lambda }\left(
-2/\varepsilon _{1},-\left( s_{0}+\delta \right) /\varepsilon _{1}\right)
\psi _{g}\left( -\left( s_{0}+\delta \right) /\varepsilon _{1}\right)
\right\| >1-o\left( \varepsilon _{1}\right)   \label{in1}
\end{equation}
and we will choose $\phi _{\varepsilon _{1}}^{\prime }=U_{\varepsilon
,\lambda }\left( -2/\varepsilon _{1},-\left( s_{0}+\delta \right)
/\varepsilon _{1}\right) \psi _{g}\left( -\left( s_{0}+\delta \right)
/\varepsilon _{1}\right) $, where the index $\varepsilon _{1}$ emphasizes
that this vector depends only on $\varepsilon _{1}$. Again, from the
adiabatic theorem on $\left( -\left( s_{0}+\delta \right) /\varepsilon
_{1},0\right) $ we have 
\begin{equation}
\left\| P_{H_{\lambda }}\left( \sigma _{-}\right) U_{\varepsilon ,\lambda
}\left( 0,-2/\varepsilon _{1}\right) \phi _{\varepsilon _{1}}^{\prime
}\right\| >1-o\left( \varepsilon _{1}\right) \text{.}  \label{in2}
\end{equation}
Because $W_{\lambda }^{\pm }$ are complete, there exists $\tilde{\phi}%
_{\varepsilon _{1}}\in P_{H_{0}}\left( \sigma _{-}\right) {\cal H}$, $%
\left\| \tilde{\phi}_{\varepsilon _{1}}\right\| \leq 1$, such that: 
\begin{equation}
W_{\lambda }^{+}\tilde{\phi}_{\varepsilon _{1}}=P_{H_{\lambda }}\left(
\sigma _{-}\right) U_{\varepsilon ,\lambda }\left( 0,-2/\varepsilon
_{1}\right) \phi _{\varepsilon _{1}}^{\prime }\text{.}
\end{equation}
In fact, $\tilde{\varphi}_{\varepsilon _{1}}$ is given by: 
\begin{equation}
\tilde{\phi}_{\varepsilon _{1}}=P_{H_{0}}\left( \sigma _{-}\right) \left(
W_{\lambda }^{+}\right) ^{\dagger }U_{\varepsilon ,\lambda }\left(
0,-2/\varepsilon _{1}\right) \phi _{\varepsilon _{1}}^{\prime }\text{.}
\end{equation}
Thus we can continue: 
\begin{equation}
\left\| P_{H_{0}}\left( \sigma _{-}\right) e^{iH_{0}2/\varepsilon
_{2}}U_{\varepsilon ,\lambda }\left( 2/\varepsilon _{2},0\right)
U_{\varepsilon ,\lambda }\left( 0,-2/\varepsilon _{1}\right) \phi
_{\varepsilon _{1}}^{\prime }\right\|   \label{in3}
\end{equation}
\[
\geq \left| \left\langle \tilde{\phi}_{\varepsilon
_{1}},e^{iH_{0}2/\varepsilon _{2}}U_{\varepsilon ,\lambda }\left(
2/\varepsilon _{2},0\right) U_{\varepsilon }\left( 0,-2/\varepsilon
_{1}\right) \phi _{\varepsilon _{1}}^{\prime }\right\rangle \right| 
\]
\[
\geq \left| \left\langle W_{\lambda }^{+}\,\tilde{\phi}_{\varepsilon
_{1}},U_{\varepsilon ,\lambda }\left( 0,-2/\varepsilon _{1}\right) \phi
_{\varepsilon _{1}}^{\prime }\right\rangle \right| 
\]
\[
-\left| \left\langle \left[ U_{\varepsilon ,\lambda }^{\ast }\left(
2/\varepsilon _{2},0\right) e^{-iH_{0}2/\varepsilon _{2}}-W_{\lambda }^{+}%
\right] \tilde{\phi}_{\varepsilon _{1}},U_{\varepsilon ,\lambda }\left(
0,-2/\varepsilon _{1}\right) \phi _{\varepsilon _{1}}^{\prime }\right\rangle
\right| 
\]
\[
=\left\| P_{H_{\lambda }}\left( \sigma _{-}\right) U_{\varepsilon ,\lambda
}\left( 0,-2/\varepsilon _{1}\right) \phi _{\varepsilon _{1}}^{\prime
}\right\| ^{2}
\]
\[
-\left| \left\langle \left[ W_{\varepsilon _{2},\lambda }^{+}-W_{\lambda
}^{+}\right] \tilde{\phi}_{\varepsilon _{1}},U_{\varepsilon ,\lambda }\left(
0,-2/\varepsilon _{1}\right) \phi _{\varepsilon _{1}}^{\prime }\right\rangle
\right| 
\]
\[
>1-o\left( \varepsilon _{1}\right) -\left| \left\langle \left[
W_{\varepsilon _{2},\lambda }^{+}-W_{\lambda }^{+}\right] \tilde{\phi}%
_{\varepsilon _{1}},U_{\varepsilon ,\lambda }\left( 0,-2/\varepsilon
_{1}\right) \phi _{\varepsilon _{1}}^{\prime }\right\rangle \right| \text{,}
\]
by using inequality (\ref{in2}). Finally, choosing $\phi
=e^{-iH_{0}2/\varepsilon _{1}}\phi _{\varepsilon _{1}}^{\prime }$ it follows
from (\ref{in1}) that: 
\[
\left\| P_{H_{0}}\left( \sigma _{-}\right) S_{\varepsilon ,\lambda
}P_{H_{0}}\left( \sigma _{+}\right) \phi \right\| 
\]
\begin{equation}
>\left\| P_{H_{0}}\left( \sigma _{-}\right) e^{iH_{0}2/\varepsilon
_{2}}U_{\varepsilon ,\lambda }\left( 2/\varepsilon _{2},-2/\varepsilon
_{1}\right) \phi _{\varepsilon _{1}}^{\prime }\right\| -o\left( \varepsilon
_{1}\right) \text{.}
\end{equation}
Further, from inequality (\ref{in3}) 
\[
\left\| P_{H_{0}}\left( \sigma _{-}\right) S_{\varepsilon ,\lambda
}P_{H_{0}}\left( \sigma _{+}\right) \phi _{\varepsilon _{1}}\right\| 
\]
\begin{equation}
\geq 1-o\left( \varepsilon _{1}\right) -\left| \left\langle \left[
W_{\varepsilon _{2},\lambda }^{+}-W_{\lambda }^{+}\right] \tilde{\phi}%
_{\varepsilon _{1}},U_{\varepsilon ,\lambda }\left( 0,-2/\varepsilon
_{1}\right) \phi _{\varepsilon _{1}}^{\prime }\right\rangle \right| \text{.}
\end{equation}
Because $\tilde{\phi}_{\varepsilon _{1}}$ do not depend on $\varepsilon _{2}$%
, the statement of the theorem follows from the strong convergence of the
adiabatic Moller operator to the static Moller operator.$\blacksquare $

\noindent Following \cite{BE}, one can second quantize our problem by
considering $P_{H_{0}}\left( \sigma _{\pm }\right) $ as the spaces of
particles and antiparticles (holes). If $S_{\varepsilon ,\lambda }$ can be
implemented in the Fock space, then one can follow the method of \cite{NE2}
to show that this result is equivalent with the spontaneous pair creations.

\noindent We want to point out that the local completeness of Moller
operators is essentially in the proof of the above theorem. Supposing that
they are only weakly local complete (i.e. $Ran\,P_{H_{\lambda }}\left(
\sigma _{-}\right) W_{\lambda }^{-}=Ran\,P_{H_{\lambda }}\left( \sigma
_{-}\right) W_{\lambda }^{+}\neq P_{H_{\lambda }}\left( \sigma _{-}\right)
H_{a.c.}\left( H_{\lambda }\right) $), then the eigenvector $\psi _{g}\left(
-\left( s_{0}+\delta \right) /\varepsilon _{1}\right) $ may be trapped in $%
P_{H_{\lambda }}\left( \sigma _{-}\right) \left[ Ran\,W_{\lambda }^{+}\right]
^{\perp }$ under the evolution $U_{\varepsilon }$. Unfortunately, it follows
from \cite{Ya2} that this is not a rare case. Moreover, because of infinite
dimensionality of this subspace, the weak convergence: 
\begin{equation}
w-%
\mathrel{\mathop{lim}\limits_{\varepsilon _{1}\rightarrow 0}}%
P_{H_{\lambda }}\left( \sigma _{-}\right) U_{\varepsilon ,\lambda }\left(
0,-1/\varepsilon _{1}\right) =0
\end{equation}
cannot be used to show that the vector escapes from $P_{H_{\lambda }}\left(
\sigma _{-}\right) \left[ Ran\,W_{\lambda }^{+}\right] ^{\perp }$ after a
long period of time. The conclusion is that during the ''switch on''
process, the eigenvector is most likely trapped and stays in $P_{H_{\lambda
}}\left( \sigma _{-}\right) \left[ Ran\,W_{\lambda }^{+}\right] ^{\perp }$.
Then there is no way of defining a vector similar to $\tilde{\phi}%
_{\varepsilon _{1}}$ so the above proof cannot be applied. Because $\left(
W_{\varepsilon ,\lambda }^{-}\right) ^{\dagger }$ converges only weakly to $%
\left( W_{\lambda }^{-}\right) ^{\dagger }$, there is no direct argument
against the possibility that the ''switch off'' process to bring this vector
back to $P_{H_{0}}\left( \sigma _{+}\right) {\cal H}$.

\section{Conclusions}

\noindent The last observation shows that even in this simplified form, the
problem of spontaneous transitions is not trivial. A deep question about the
subject is under what conditions the same result is true disregarding any
order of the limits, in particular, for $\varepsilon _{1}=\varepsilon _{2}$.
In the case when Moller operators are complete (or locally complete on $%
\sigma _{-}$), the result of the last section reduces this problem to the
study of $\tilde{\phi}_{\varepsilon _{1}}$ properties. One might expect that 
\begin{equation}
\int_{0}^{\infty }dt\left\| Ve^{-itH_{0}}\tilde{\phi}_{\varepsilon
_{1}}\right\| <M\text{,}  \label{last}
\end{equation}
with $M$ independent of $\varepsilon _{1}$ in which case it is
straightforward that the order of limits is unimportant. To prove a relation
like \ref{last}, one has to prove that $\tilde{\phi}_{\varepsilon _{1}}$
belongs to a set of vectors for which the Cook criterion is valid, together
with uniform estimates. From the definition of $\tilde{\phi}_{\varepsilon
_{1}}$, one can see that this problem can be reduced to the study of the
evolution of the eigenvector $\psi \left( -\left( s_{0}+\delta \right)
/\varepsilon _{1}\right) $, which does not depend on $\varepsilon _{1}$. In
the most of the cases, the Schwartz's space may be chosen as the set of
vectors for which the Cook criterion holds. Unfortunately, to prove that the
evolution of $\psi \left( -\left( s_{0}+\delta \right) /\varepsilon
_{1}\right) $ belongs to this space is almost impossible. A much easier task
is to prove that it belongs to some Sobolev space $W^{k,p}$. If this step is
accomplished, we think that $W^{k,p}$ estimates of \cite{Yaj} may be used to
complete the proof, at least for large dimensions.

\begin{acknowledgement}
The author gratefully acknowledges support, under the direction of J.
Miller, by the State of Texas through the Texas Center for Superconductivity
and the Texas Higher Education Coordinating Board Advanced Technology
Program, and by the Robert A. Welch Foundation. Also, the author acknowledge
extremely helpful observations pointed out by one of the referees.
\end{acknowledgement}

\end{document}